comp-gas/9501009 23 Jan 1995

# Plane Waves in a Multispeed Discrete-Velocity Gas

B. T. Nadiga
California Institute of Technology
Pasadena, California 91125

## Abstract

A kinetic flux-splitting procedure used in conjunction with local thermodynamic equilibrium in a finite volume allows us to investigate numerically discrete-velocity gas flows. The procedure, outlined for a general discrete-velocity gas, is used to simulate flows of the nine-velocity gas, a simple two dimensional multiple-speed discrete-velocity gas, wherein a multiplicity of speeds ensures nontrivial thermodynamics. After verifying the linear wave limit and the non-linear steepening of wavefronts, the stability and propagation of planar discontinuities in that model gas is studied. The supersonic-subsonic requirement for the stable propagation of a discontinuity, being kinematic in nature is the same in the model gas, as *e.g.*, in a perfect gas. However, the finiteness of the velocity space in the model gas does not allow a translation of the above kinematic condition to the thermodynamic requirement of increasing entropy across a compressive shock: a case of an entropy decreasing compressive shock in the model gas is presented. Finally, the interaction of various types of waves—shock waves, rarefactions and contact surfaces—in the model gas are shown in a simulation of the shock tube problem.

## Introduction

Regular discretization of the velocity space of particles constituting a gas results in a discrete-velocity gas. The computational evolution of such a system, especially the collision phase, is greatly simplified if the discretization is not elaborate. Lattice gases involve a discretization of the physical space as well, allowing them to be mapped directly on to a digital computer, resulting in elegant computational models for fluids. The work by Frisch et. al.[1], wherein the incompressible Navier-Stokes equations were recovered for the Frisch-Hasslacher-Pomeau (FHP) model, and the subsequent success of using lattice gases[2,3] in simulating various fluid phenomena have renewed interest in these discrete-velocity models, first used in a flow situation by Broadwell[4,5]. The predominant use of single-speed models to simulate incompressible flow notwithstanding, the macroscopic behavior of these models is compressible. Therefore, to investigate the applicability of discrete-velocity models to compressible flows, we have been studying a few multiple-speed models; the independent energy variable in the multiple-speed models allows nontrivial thermodynamics. We note that compressible flow dynamics of single speed models have been studied[5–7].

17 June 1993

The computational technique[8] with the attendant flux-splitting and total-variation-diminution[9] ideas are outlined after briefly describing the nine-velocity gas[10−13]. Linear wave propagation and non-linear steepening and shallowing of planar waves—a familiar consequence of the convective nonlinearity in fluid equations—are then considered. Next, jump conditions associated with the model gas, time evolution of the jumps and how they relate to thermodynamics of a perfect gas are studied. We end with a simulation of the shock-tube problem in the model gas, and a few conclusions.

### The Nine-Velocity Gas

The nine-velocity gas, a two-dimensional model, consists of a large number of identical hard sphere (disk) particles, each of which takes on one of the nine allowed velocities

$$\begin{array}{lll} \mathbf{c}_0=(0,0), & \mathbf{c}_1=(1,0), & \mathbf{c}_2=(1,1) \\ \mathbf{c}_3=(0,1), & \mathbf{c}_4=(-1,1), & \mathbf{c}_5=(-1,0) \\ \mathbf{c}_6=(-1,-1), & \mathbf{c}_7=(0,-1), & \mathbf{c}_8=(1,-1) \end{array}$$

Except for the simplification of discretizing the velocity space, the model gas is much like a monatomic gas with a hard-sphere potential, and consists of free-flight interrupted by instantaneous collisions with other particles. The collisions, in addition to conserving mass, momentum, and energy individually, are closed under the velocity set. The four types of collisions possible are

$$\begin{array}{ll} \text{type 1: } (\mathbf{c}_3,\, \mathbf{c}_7) \rightarrow (\mathbf{c}_1,\, \mathbf{c}_5), & \text{type 2: } (\mathbf{c}_4,\, \mathbf{c}_8) \rightarrow (\mathbf{c}_2,\, \mathbf{c}_6) \\ \text{type 3: } (\mathbf{c}_1,\, \mathbf{c}_3) \rightarrow (\mathbf{c}_0,\, \mathbf{c}_2), & \text{type 4: } (\mathbf{c}_1,\, \mathbf{c}_6) \rightarrow (\mathbf{c}_5,\, \mathbf{c}_8) \end{array}$$

Collision type 3 is unique in that the precollision speeds are different from the postcollision speeds, and this provides the crucial mechanism for equilibration between the various particle speeds.

If $\mathbf{n}$ $[n_i=n(\mathbf{c}_i),\ i = 0, 1, \ldots, 8]$ represents the full velocity distribution function, then at thermodynamic equilibrium, from a detailed balancing of collisions, there are five relations among the nine population densities:

$$\begin{aligned} &n_1n_3 = n_0n_2, \quad n_3n_5 = n_0n_4, \quad n_5n_7 = n_0n_6 \\ &n_7n_1 = n_0n_8, \quad n_1n_5 = n_3n_7 \end{aligned} \tag{1}$$

From Eq. (1) only four of the $n_i$ are independent; a set of four such particle populations is denoted by $\mathbf{m}$, where $\mathbf{m}$ is a Maxwellian state.

The vector of mass, momentum, and energy, denoted by $\mathbf{F}$, is given by

$$\mathbf{F} = \left( \sum_{a=0}^{8} n_a, \sum_a n_a\mathbf{c}_a, \sum_a n_a\mathbf{c}_a^2 \right) = (\rho, \rho\mathbf{u}, \rho E) \tag{2}$$

where $E = e + \frac{\mathbf{u}^2}{2}$ and $\mathbf{G}$, the flux of $\mathbf{F}$ is given by

$$\mathbf{G} = \left( \sum_a n_a\mathbf{c}_a, \sum_a n_a\mathbf{c}_a\mathbf{c}_a, \sum_a n_a\mathbf{c}_a^2\mathbf{c}_a \right) \tag{3}$$

With the above definitions of $\mathbf{F}$ and $\mathbf{G}$, and the equilibrium relations Eq. (1), the Euler equations for the model gas may be written implicitly in terms of $\mathbf{m}$ as

$$\frac{\partial \mathbf{F}(\mathbf{m})}{\partial t} + \nabla \cdot \mathbf{G}(\mathbf{m}) = 0 \tag{4}$$

Pressure in the model gas, as described above is[14]

$$p = \rho \left[ e - \frac{1-3e}{4e}\mathbf{u}^2 + O(\mathbf{u}^3) \right] \tag{5}$$

correct to $O(\mathbf{u}^2)$: Note that at $e = \frac{1}{3}$, Eq. (5), the equation of state of the model gas reduces to that of an ideal gas. Because of the kinematic dependence of the thermodynamics, the speed of propagation of sound in such a gas is dependent on the macroscopic flow velocity, and because of that, there is not a unique speed of sound to which velocities can be referred[14]. This makes the definition of a Mach number in a discrete-velocity gas neither natural nor unique.

### Simulation Scheme for a Discrete-Velocity Gas

We briefly outline the computational technique here; the details can be found in Nadiga and Pullin[8]. For simplicity, a flow in one spatial dimension is considered: the flow is along the $x$ axis, coincident with $\mathbf{c}_1$. Instead of dealing with individual particles, the flowfield is tiled with a linear array of cells, with each cell interacting with its adjacent neighbors through a flux of $\mathbf{F}$, the vector of mass, momentum, and energy. The crucial step of the technique consists of splitting $\mathbf{G}$, the flux of $\mathbf{F}$, based on the signs of the discrete-velocities of the particles into $\mathbf{G}^+$ and $\mathbf{G}^-$ and then computing the split fluxes in accordance with local thermodynamic equilibrium. The time evolution then simply consists of noting that $\mathbf{F}$ is conserved in time and updating the particle distribution function in each of the cells accordingly. To illustrate the procedure, we write a first-order scheme based on the above ideas using an Euler time step:

$$\begin{aligned} \mathbf{m}(x, t+\Delta t) &= \mathbf{m}(x,t) \\ &- \frac{\Delta t}{\Delta x}[\mathbf{J}_{\mathbf{Fm}}]^{-1} \{ \mathbf{G}^+[\mathbf{n}^+(x,t)] - \mathbf{G}^-[\mathbf{n}^-(x+\Delta x, t)] \\ &\qquad - \mathbf{G}^+[\mathbf{n}^+(x-\Delta x, t)] + \mathbf{G}^-[\mathbf{n}^-(x,t)] \} \end{aligned} \tag{6}$$

$\mathbf{J}_{\mathbf{Fm}}$ is the Jacobian of the transformation from $\mathbf{F}$ to $\mathbf{m}$, and follows from the definition of $\mathbf{F}$ in Eq. (2), $\mathbf{n}^+$ is the distribution function of particles with a positive $u$-velocity component and $\mathbf{n}^-$ that of particles with a negative $u$-velocity component. According to Eq. (6), the distribution function at $x$ at time $t+\Delta t$ is different from that at time $t$ by 1) the departure of particles from $x$ due to a non-zero $u$-velocity, terms 1 and 4; 2) the arrival of particles with a positive $u$-velocity from $x-\Delta x$, term 3; and 3) the arrival of particles with a negative $u$-velocity from $x+\Delta x$, term 2. However, the important difference from the usual discrete-velocity gas evolution is the fact that the arrival and departure of the particles is in accordance with the equilibrium split-flux of $\mathbf{F}$. In the present case of the model gas, with the flow along $\mathbf{c}_1$, the definition of $\mathbf{G}$ used in conjunction with Eq. (1) gives the expressions for the local equilibrium fluxes, and while particles with velocities $\mathbf{c}_1$, $\mathbf{c}_2$, and $\mathbf{c}_8$ contribute to $\mathbf{G}^+$, particle types $\mathbf{c}_4$, $\mathbf{c}_5$, and $\mathbf{c}_6$ contribute to $\mathbf{G}^-$.

The minmod[9] limiter is defined as the binary operator:

$$\mathrm{minmod}(p,q) = \mathrm{sgn}(p)\begin{cases} 0 & \text{if } \mathrm{sgn}(p) \neq \mathrm{sgn}(q) \\ \min\{|p|,|q|\} & \text{if } \mathrm{sgn}(p) = \mathrm{sgn}(q) \end{cases} \tag{7}$$

where $\mathrm{sgn}(p)$ is the sign of $p$ and $|p|$ is the absolute value of $p$. In the present usage, $p$ and $q$ are the values of the slopes at the centroid of a cell: $p$ being the backward slope and $q$ the forward slope. The full minmod limiting procedure simply consists of applying the above binary operation to each of the cells at any given time step to obtain the slopes for any relevant quantity at each of the cell centroids. Using the minmod limiter to interpolate $\mathbf{F}$ gives rise to slope-limited schemes, whereas using it to interpolate $\mathbf{G}$ gives rise to flux-limited schemes. In either case, the minmod limiting strategy results in a diminution of the total variance of the interpolated field, giving a stable second-order scheme. There being no inherent advantage of either the flux-limited or the slope-limited scheme over one another, we use the flux-limited scheme. The Euler time stepping in Eq. (6) was purely illustrative; we use a fourth order Runge-Kutta scheme in all the cases.

## Linear Wave Equation Limit

The initial disturbance, imposed over an uniform state ($\rho_0$=1, $e_0$=0.33, $u_0$=0.00) of the model gas in a periodic [0,1] domain in the direction of $\mathbf{c}_1$, is in the form of a hyperbolic secant variation in density. The time evolution of pressure, as defined in Eq. (5), is shown in Fig. 1. The behavior of the other quantities is identical, in concurrence with linearity of the (nondispersive) wave behavior. Throughout the computation, the values of mass, momentum, and energy integrated over the domain are all conserved to better than one part in a million. On reversing time, the two waves recombine to give the initial disturbance correct to the irreversible diffusion arising out of the numerical viscosity[12] of the technique, but which can be interpreted in terms of the viscosity of the model gas[14] (and imperfections in the initial data).

## Nonlinear Wave Steepening

The best setup for this is of course a *simple* compression wave—only one of the three families of characteristics not parallel—propagating into an uniform state. Not knowing the Riemann invariants[14] in the model gas prevents such an explicit setup. Therefore, the time evolution of an initial hyperbolic tangent interface between two equilibria (which satisfy the Rankine-Hugoniot jump conditions[15] arising from Eq. (4)) is studied. The gradients of all of the flow variables at the inlet and the outlet are prescribed to be zero. Figure 2 compares the steepness of the pressure profiles. The steepening of the front is clearly indicative of the nonlinearity: the speed of propagation of the disturbance $\omega_+$ varies as the magnitude of the disturbance (say $p$) [cf. $\mathrm{d}(u+a) = \frac{\Gamma}{\rho c}\mathrm{d}p$ for a perfect gas]. Figure 2 also shows how an initially steep expansion front gets shallower with time. (The situation in neither of the above two cases is *exactly* that of a simple wave because all that can be said is that the variation spans across two uniform regions.)

## Jump Conditions

Although all of the discussion in this paper assumes the frame of reference of the model, to write an expression for the jump across an infinitesimally thin discontinuity, occuring in Eq. (4) and moving with a velocity $U$ in the direction of $\mathbf{c}_1$, we momentarily fix the frame of reference with the discontinuity; then $\partial/\partial t$ can be replaced by $-U\partial/\partial x$ and integrating the resulting equations from ahead of the shock to behind it gives the shock jump conditions[15]

$$[\![\mathbf{G}(\mathbf{m})]\!] = U[\![\mathbf{F}(\mathbf{m})]\!] \quad \text{or equivalently} \quad [\![\mathbf{G}(\mathbf{n})]\!] = U[\![\mathbf{F}(\mathbf{n})]\!] \tag{8}$$

where $[\![x]\!] = x_d - x_u$.

## Evolution of Planar Discontinuities

The computational setup used in studying the discontinuities is exactly like the one used for studying the steepening of a compressive wave: the flux-limited, second-order scheme in the domain [0,1] with Neumann boundary conditions at inflow and outflow. The initial conditions in all the cases is a step discontinuity $\mathbf{J}_0$ between the upstream state and the downstream state

$$\mathbf{J}_0 : (\rho_u, e_u, u_u) \xrightarrow{U} (\rho_d, e_d, u_d) \tag{9}$$

obtained as a solution of the algebraic system Eq. (8), unless stated otherwise. The subsequent time evolution of the discontinuity is studied. All of the quantities in this paper are in their nondimensional form: density corresponds to the average number of particles in a cell, velocities are nondimensionalized by $q$, the unit speed of the model, and energies by $q^2$.

**Case 1**

The initial condition in Fig. 3

$$\mathbf{J}_0 : (1.000, 0.100, 0.000) \xrightarrow{U} (1.137, 0.152, 0.100) \tag{10}$$

has no explicit information about the direction of propagation of the discontinuity $\mathbf{J}_0$ or its speed. $\mathbf{J}_0$ satisfies the jump conditions of the Euler equations, Eq. (8), for $U$=0.828. Evolving the initial conditions, the jump $\mathbf{J}_0$ is seen to be preserved with time and it propagates with a speed $U$=0.828 to the right, exactly as obtained from the model Euler equations. The propagating discontinuity is compressive: this follows from a calculation of the pressure given by Eq. (5) (the concurrence of the above definition of pressure with its mechanical counterpart has to be looked into) upstream and downstream. The entropy, defined using a Boltzmann-like H function as

$$s = -\frac{1}{\rho}\sum_{0}^{8} n_i \ln(n_i) \tag{11}$$

is seen to rise across the shock in this case. The characteristic velocities upstream and downstream are $C_u = (\mathbf{0.742}, -0.742, 0.000)$ and $C_d = (\mathbf{0.881}, -0.554, 0.014)$, where they are expressed as $(\omega_+, \omega_-, \omega_0)$, the three characteristics of Eq. (4).

**Case 2**

Time evolution of the initial jump

$$\mathbf{J}_0 : (1.000, 0.200, 0.000) \xrightarrow{U} (0.872, 0.141, -0.100) \tag{12}$$

shows that the jump is stable in the sense that it does not break up into different jumps, but it is seen to disintegrate with time: the slopes of all the quantities decrease continuously with time. The jump conditions, Eq. (8), do not rule out such discontinuities; and, therefore, a possible external constraint has to be invoked. Kinematically, this is seen to be the supersonic-subsonic condition: the discontinuity must travel faster than the speed of propagation of an infinitesimal disturbance in the direction of propagation of the discontinuity upstream and slower than the speed of propagation of a disturbance in the same direction downstream. Thus, for a $C_+$ discontinuity, $U$ should be such that $(\omega_+)_u \leq U \leq (\omega_+)_d$. We note parenthetically that whereas a similar relation would hold for a $C_-$ shock (but with the absolute values of the speeds concerned), a $C_0$ family of shocks are also possible by replacing subscript + with subscript 0. A discontinuity thus propagates stably if the effect of the passage of the discontinuity at a point is to increase the speed of propagation of a small disturbance in the direction concerned there; for a stable $C_+$ discontinuity, $[\![\omega_+]\!] > 0$. The disintegration of the jump $\mathbf{J}_0$, Eq. (12), which satisfies the jump condition is then explicable as due to the violation of the supersonic-subsonic condition. The characteristic velocities upstream and downstream are $C_u = (\mathbf{0.775}, -0.775, 0.000)$ and $C_d = (\mathbf{0.531}, -0.886, -0.013)$, and the initial shock velocity $U = \mathbf{0.680}$.

$$(\omega_+)_u \not\leq U \not\leq (\omega_+)_d \quad \text{or} \quad [\![\omega_+]\!] \not> 0 \tag{13}$$

The stable shock in case 1 is seen to satisfy the supersonic-subsonic condition.

**Case 3**

$$\mathbf{J}_0 : (1.000, 0.400, 0.000) \xrightarrow{U} (1.292, 0.455, 0.200) \tag{14}$$

satisfies the jump conditions, Eq. (8), for $U$=0.885. Evolving the initial condition, $\mathbf{J}_0$ is seen to be preserved with time, see Fig. 4, propagating with a speed $U$=0.885 to the right, again in concurrence with the Euler equations. The main difference between the present case and case 1 is that the entropy of the downstream is lower than that of the upstream. If the definition of entropy used is correct, the behavior of entropy in the present case may appear counter to what is expected from the second law of thermodynamics. To better understand this, we consider the behavior of entropy in the more familiar case of a perfect gas. The kinematic supersonic-subsonic requirement, $[\![u+a]\!] > 0$, (more generally valid for any normal[16] fluid), can be related to the variation of thermodynamic quantities across the shock by

$$\mathrm{d}(u+a) = \frac{\Gamma}{\rho a}\mathrm{d}p. \tag{15}$$

where $\Gamma$ is the fundamental derivative of gasdynamics $\frac{1}{a}\left[\frac{\partial \rho a}{\partial \rho}\right]_s$, $\rho a$ the acoustic impedance and $a$ the speed of sound. Therefore, across a shock,

$$[\![u+a]\!] = \int_{p_u}^{p_d} \frac{\Gamma}{\rho a}\mathrm{d}p \tag{16}$$

With $\Gamma = \frac{\gamma+1}{2} > 0$ for a perfect gas, the supersonic-subsonic requirement can be satisfied only with $p_d > p_u$ or through a compressive shock, and the following relation[16] ensures a positive entropy jump across a compressive shock:

$$\frac{T_u[\![s]\!]}{a_u^2} = \frac{1}{6}\Gamma_u\Pi^3 + O(\Pi^4) \qquad \text{where } \Pi = \frac{[\![p]\!]}{\rho_u a_u^2} \tag{17}$$

However, for a perfect gas, the entropy jump can be easily expressed in terms of the upstream and downstream quantities

$$\mathrm{d}s = c_p\frac{\mathrm{d}e}{e} - R\frac{\mathrm{d}p}{p} \quad \Rightarrow \quad [\![s]\!] \propto \log_e\left\{\frac{p_u}{p_d}\left(\frac{e_d}{e_u}\right)^{\frac{\gamma}{\gamma-1}}\right\}. \tag{18}$$

Since the model fluid being studied is two dimensional, $\gamma$, the ratio of specific heats, is 2. Therefore

$$[\![s]\!] \propto \log_e\left\{\frac{\rho_u}{\rho_d}\frac{e_d}{e_u}\right\} \tag{19}$$

The Rankine-Hugoniot relations across a compressive shock ($M_u > 1$) in a perfect gas are such that $\frac{\rho_u}{\rho_d}\frac{e_d}{e_u} > 1$, ensuring a positive entropy jump across it.

Since the particles in the model gas are hard spheres, like in an ideal gas, we assume the model gas to be a normal fluid, *i.e.*, $\Gamma > 0$. Although we do not have the counterpart of Eq. (15) in the model gas, relating $d(\omega_+)$ to $dp$, we point out that if the relation is similar to Eq. (15), then $\Gamma > 0$ implies the possibility of only compressive shocks, and that is what is computationally observed. We cannot use Eq. (17) in the context of the model gas because it uses conjugate variables, temperature of specific energy and pressure of density, and their definition is presently not clear due to the finiteness of the velocity space.

In Fig. 5, the entropy jump across a series of shocks in the model gas is studied. The entropy jumps are calculated by two different methods: on the $x$ axis is the entropy jump calculated using the statistical formulation, Eq. (11), and on the $y$ axis is $\log_e\left(\frac{\rho_u e_d}{\rho_d e_u}\right)$, from Eq. (19). This is done for two series of shocks: In the first series (triangles), the upstream is held constant at $(\rho_u, e_u, u_u) = (1.0, 0.4, 0.0)$ and the piston moved into it at velocities ranging from 0.01 to 0.5. In the second series (diamonds), the piston is moved at a constant velocity of 0.2 into an upstream $(\rho_u, e_u, u_u) = (1.0, 0.3 \leq e_u \leq 0.6, 0.0)$. The linear least-squares fit (the solid lines) in the two cases indicate the validity of Eq. (19) as an approximation for the model gas, or more generally the correspondence between the thermodynamics of the model gas to that of a perfect gas, in the regimes considered. Although the negativity of the entropy jump in case 3 is ascribable to the finiteness of the phase space of the model gas, it is necessary to work out the thermodynamics in detail to fully describe such behavior.

**Shock Tube Problem**

Finally, a simulation of the shock tube flow with the model gas is considered. The initial condition corresponds to a high-density, high-pressure driver section and a low-density, low-pressure driven section, separated by a diaphragm which ruptures at $t = 0$. The initial jump is

$$\mathbf{J}_0 : (1.000, 0.500, 0.000) \xrightarrow{U} (0.200, 0.200, 0.000) \tag{20}$$

The gas in the two regions is the same. The end walls are modeled by mirror image sites across their actual location, resulting in specular walls which do not permit flow through them and are adiabatic. The time evolution of density is shown using a grey-level coding on the $x$-$t$ plane in Fig. 6. The qualitative nature of the interactions of the various types of waves—the shock wave, the rarefaction fan, and the contact surface—in that picture is the same as in a perfect gas.

## Conclusions

Using a new scheme to simulate discrete-velocity gases, we were able to study the one-dimensional gas dynamics of a simple multispeed discrete-velocity gas. The scheme, based on local thermodynamic equilibrium, has a viscosity which is interpretable kinetically, due to which discontinuities are captured without any spurious oscillations. While the model gas behaves qualitatively like a perfect gas in some situations, it exhibits seemingly anamolous behavior in others. A better thermodynamic analysis of these finite phase space models is the key to understanding them.

**Fig. 1 An initial pressure pulse splits into a $C_+$ and a $C_-$ disturbance (a), each of which propagates with little distortion (parallel characteristics). Reversing time at t=0.2, the $C_+$ and the $C_-$ waves recombine to reconstruct the original initial condition (b), correct to the diffusion, verifying linear superposability of the disturbances.**

**Fig. 2 A compressive wave steepens with time (a), but a steep expansion wave becomes shallower with time (b). In both figures, the profiles have been offset by the distance they have traveled to make the comparison.**

**Fig. 3 An unsteady compressive shock in the model gas. The entropy increases across the shock.**

**Fig. 4 An entropy decreasing shock.**

**Fig. 5 A comparison of the statistical definition of entropy in the model gas to the thermodynamic definition for a 2-D perfect gas used for the model gas.**

**Fig. 6 A shock tube flow in the model gas.**

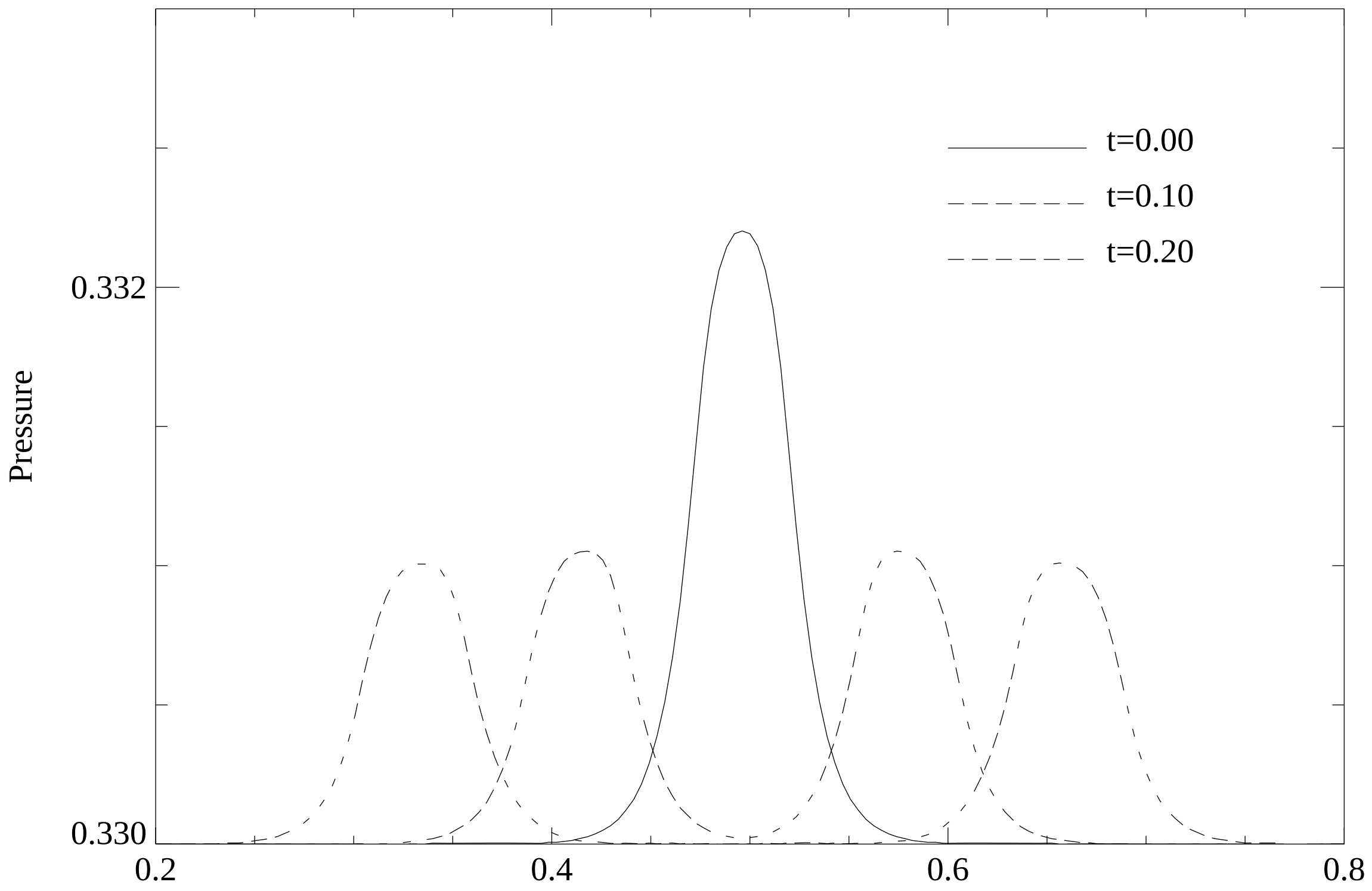

FIG. 1a

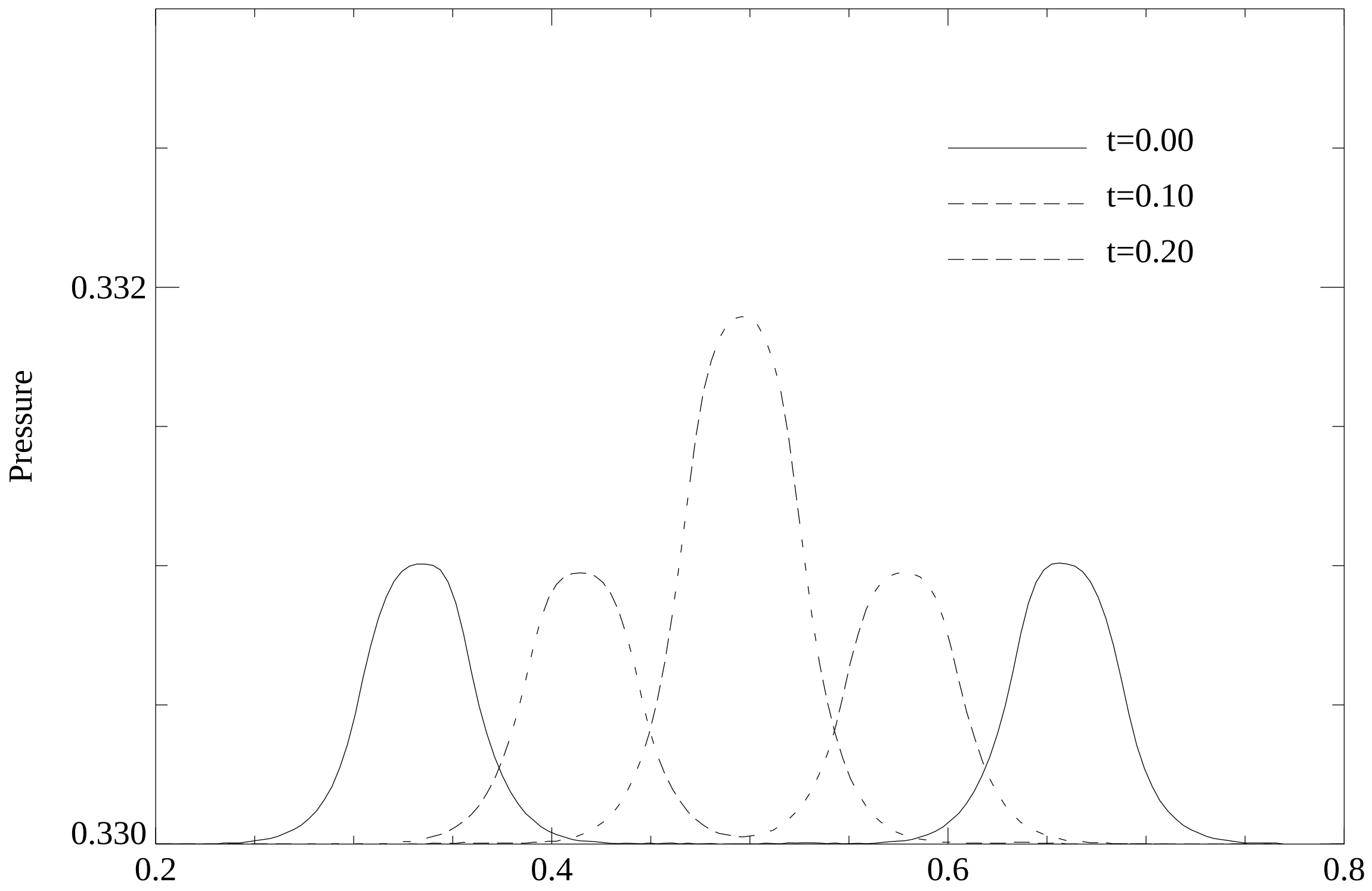

FIG. 1b

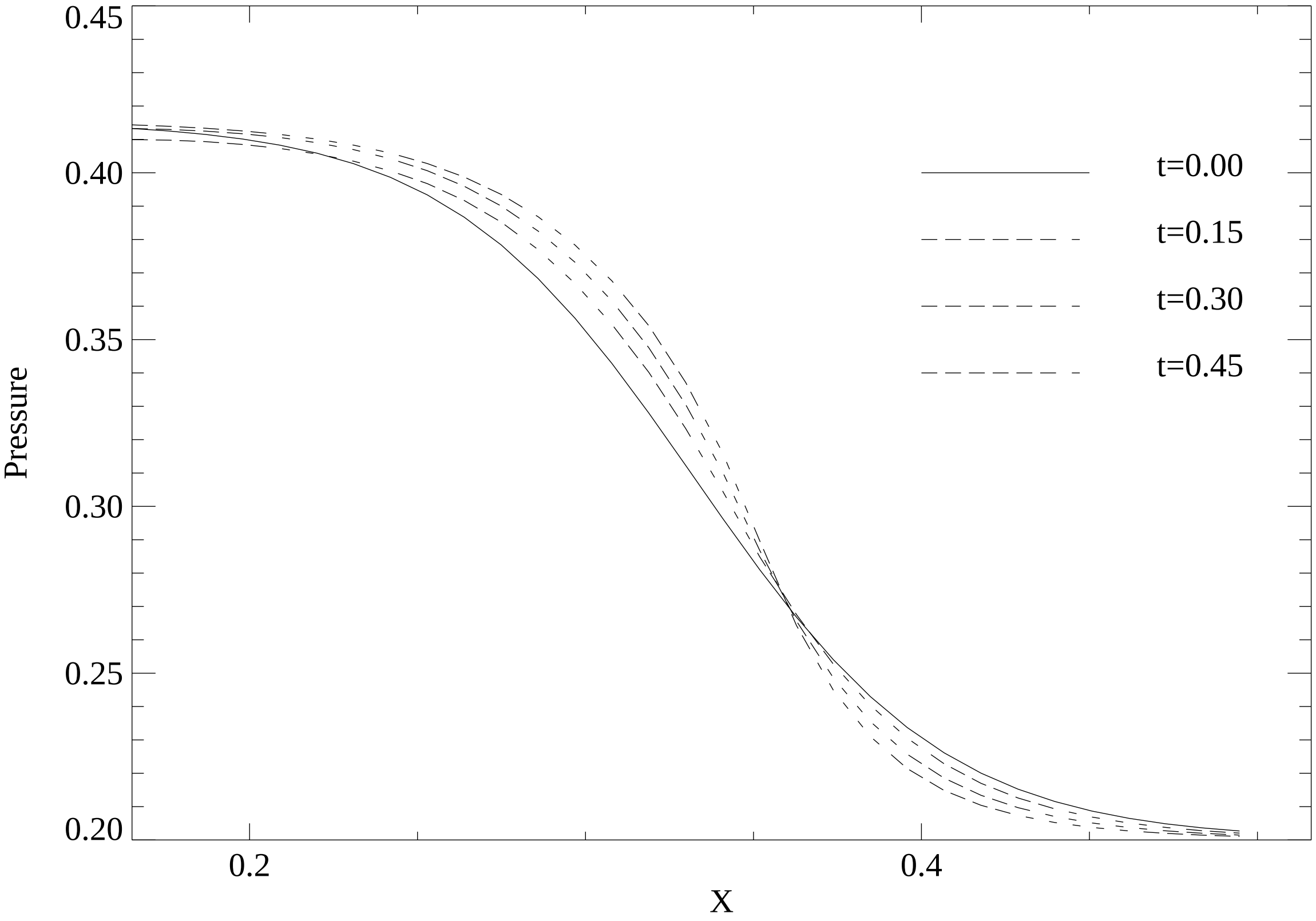

FIG. **2a**

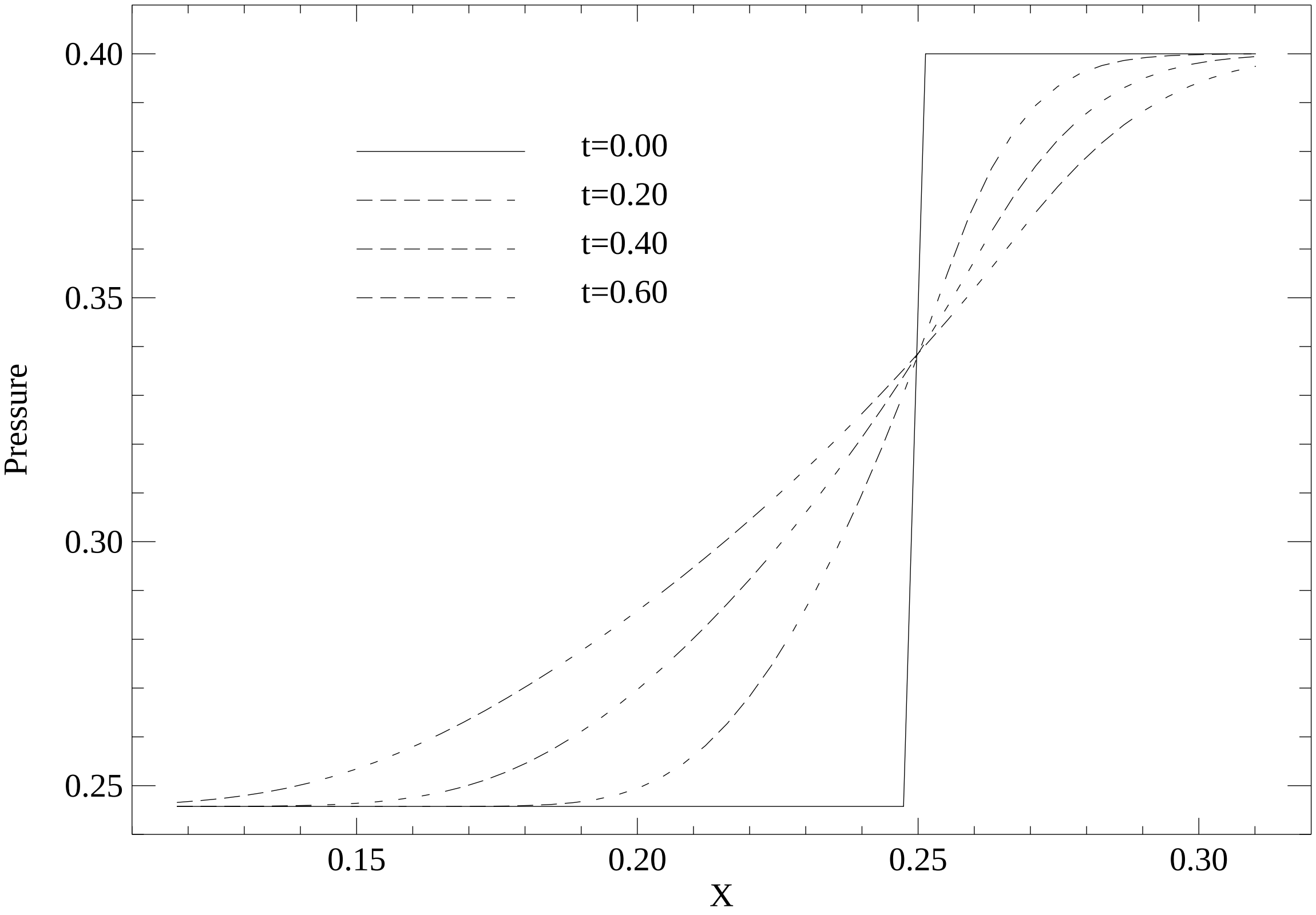

FIG. **2b**

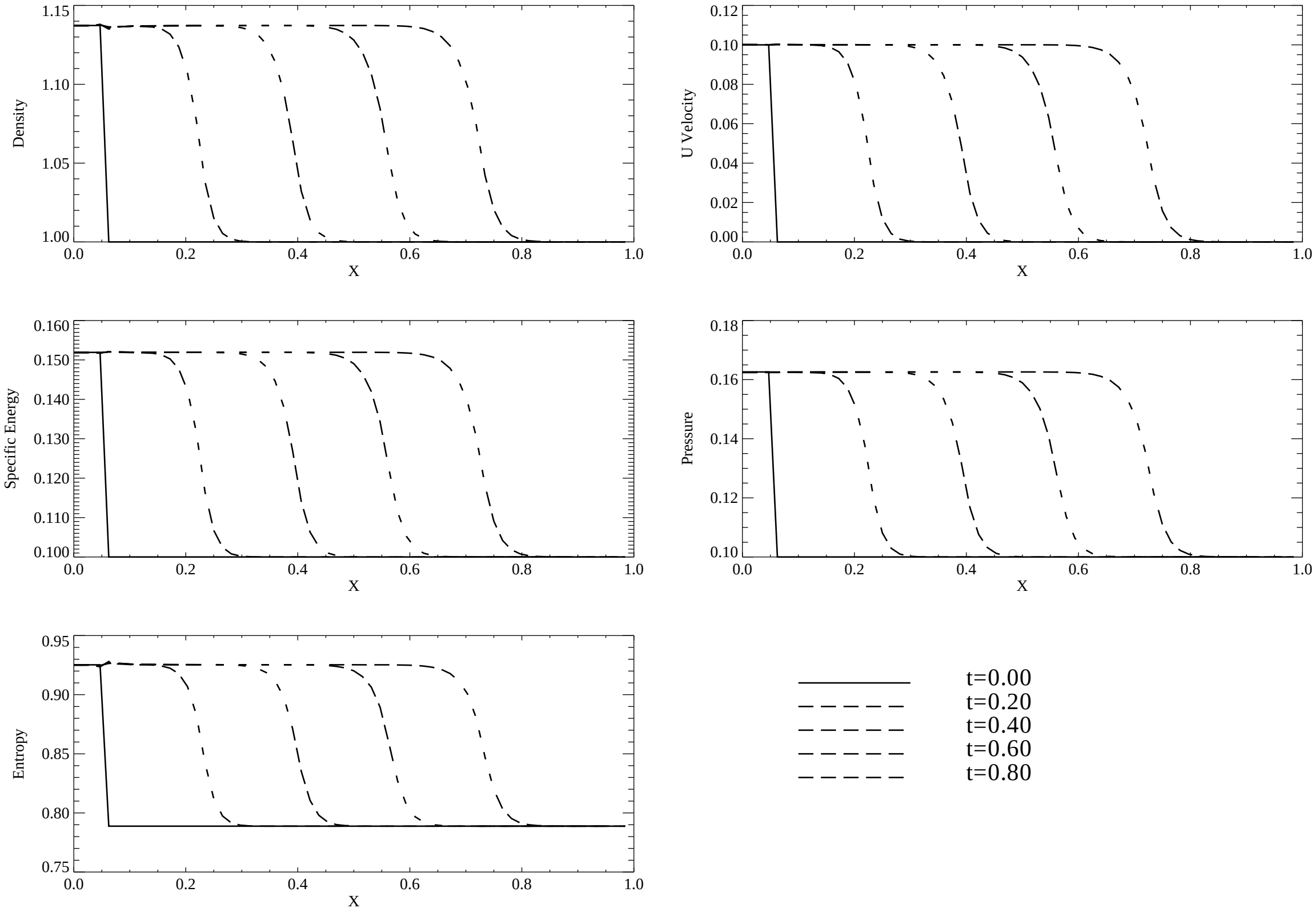

FIG. 3

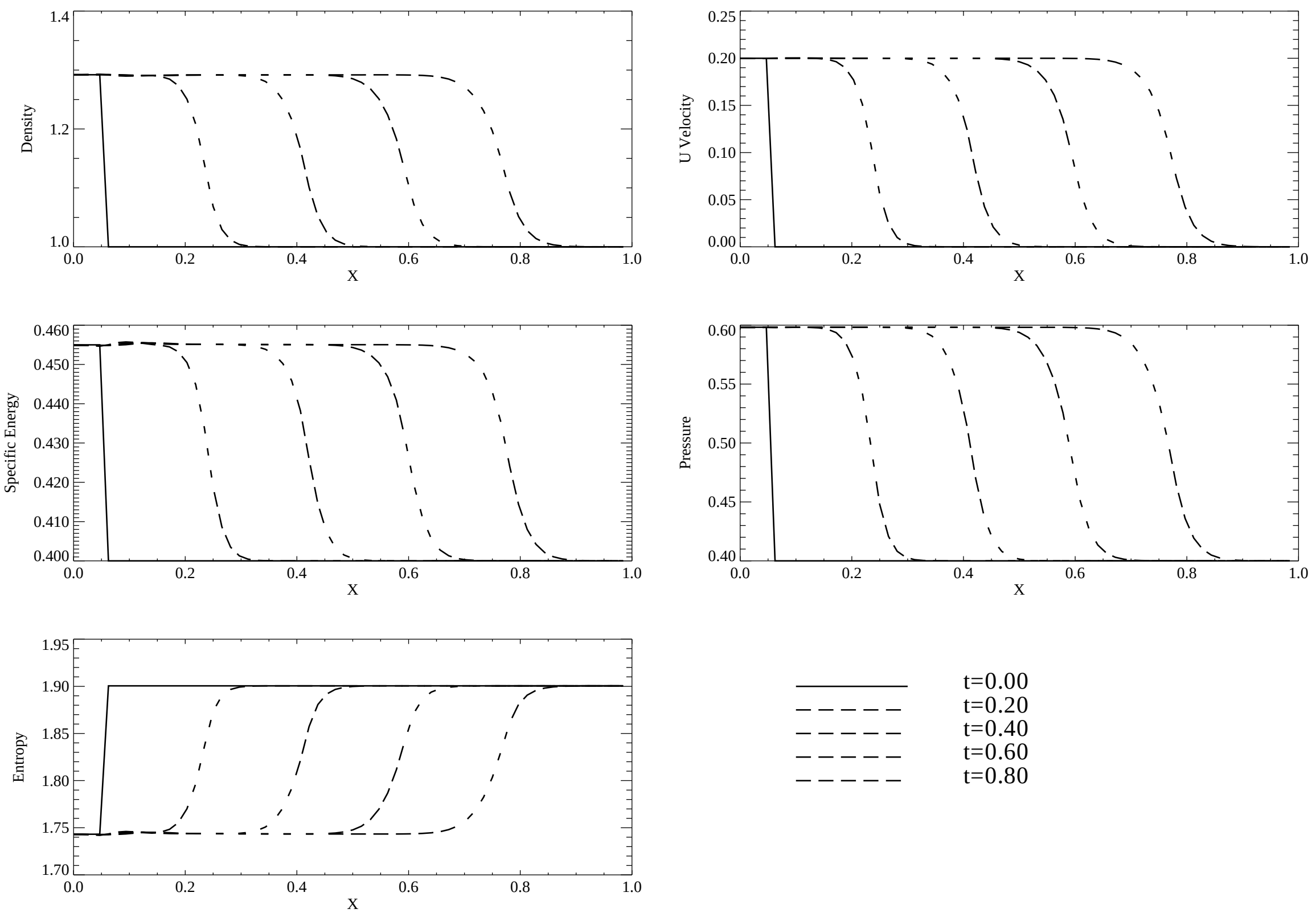

FIG. 4

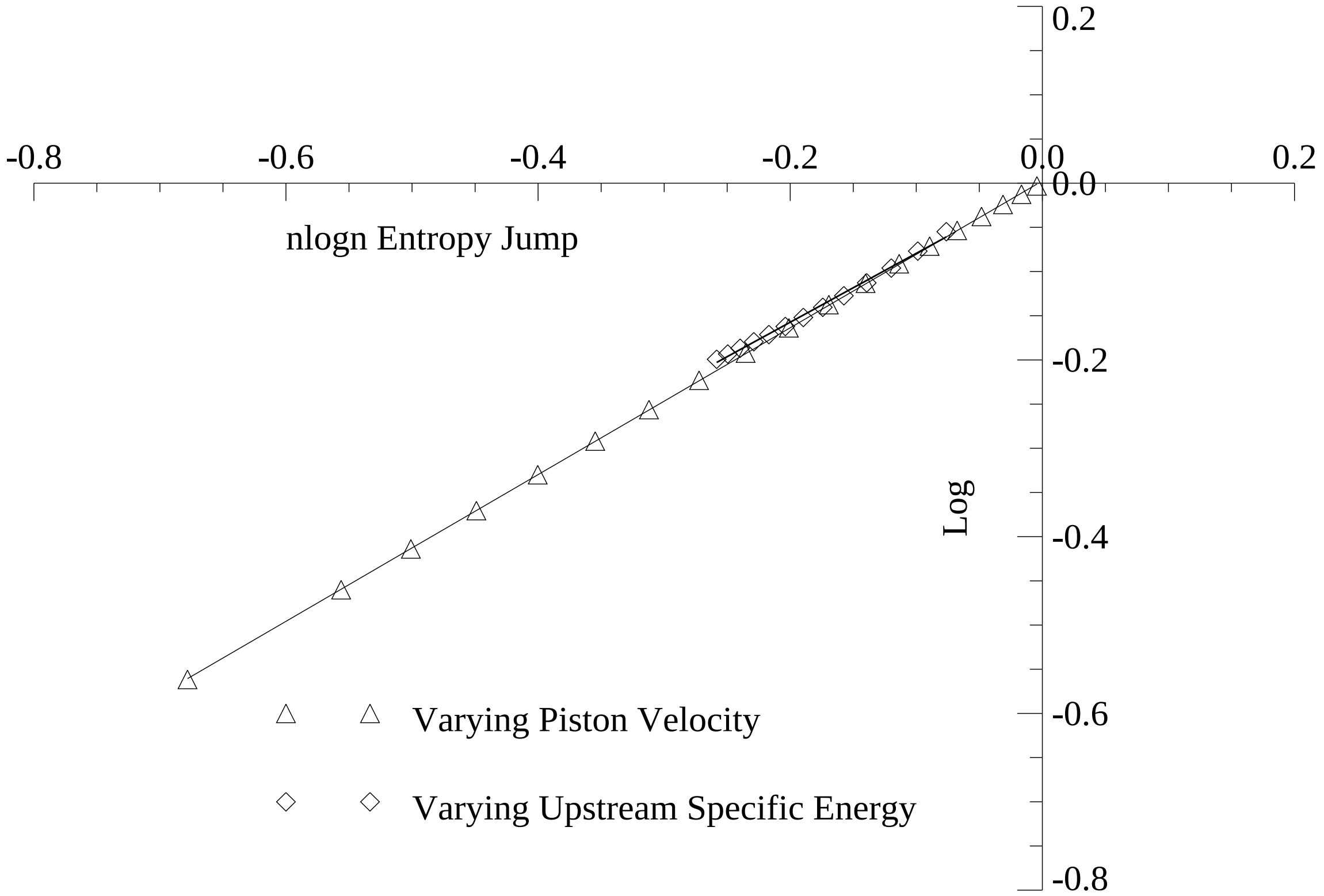

FIG. 5

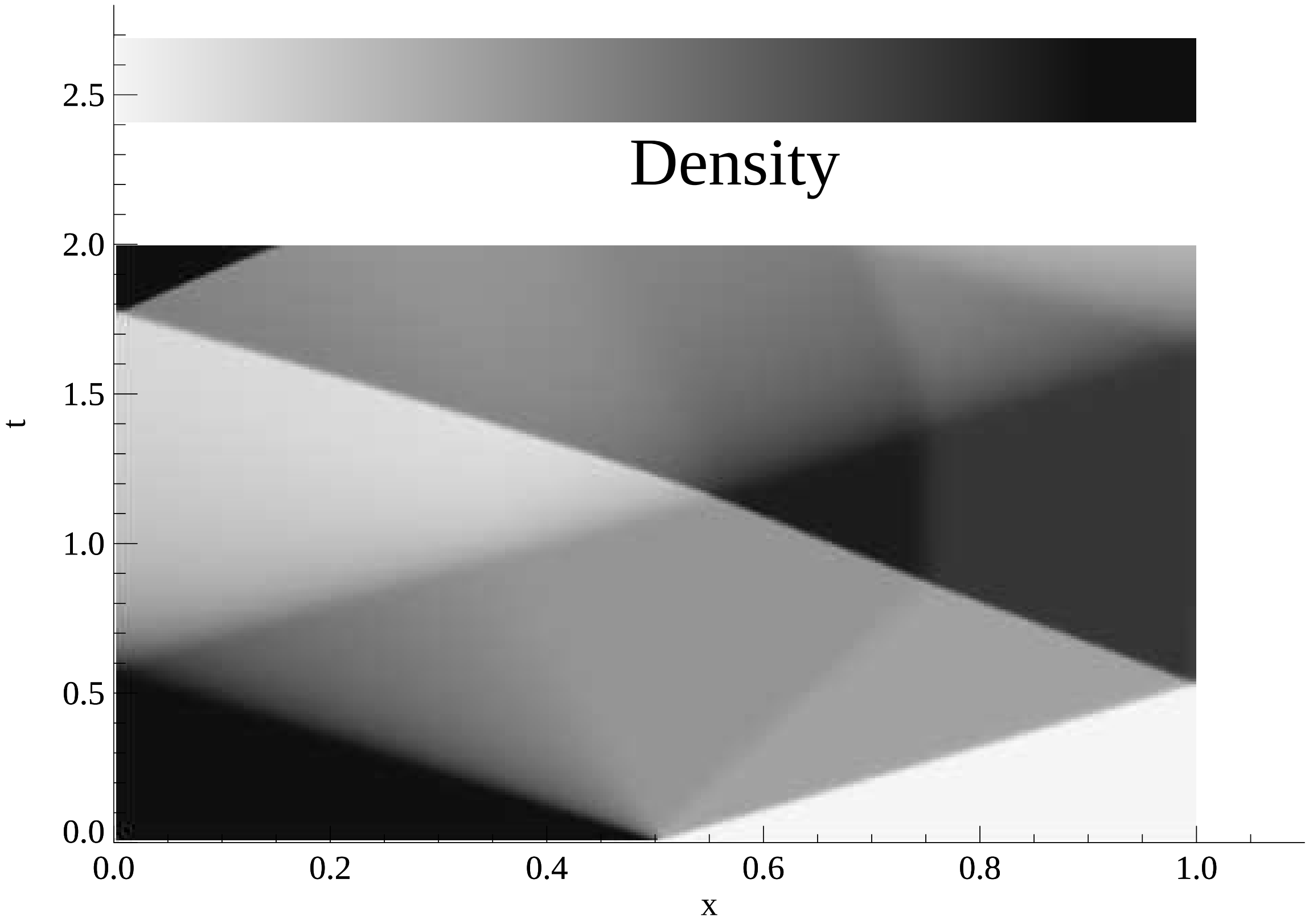

FIG. 6